\documentclass[11pt,twoside]{article}

\usepackage{asp2006}
\usepackage{epsf}
\usepackage{lscape}
\usepackage{graphicx}

\def    \bJ     {\bf J}
\def    \bB     {\bf B}

\begin{document}
\title{Quantitative Theory of Grain Alignment: Probing Grain Environment and Grain Composition}   
\author{A. Lazarian}   
\affil{Department of Astronomy, University of Wisconsin-Madison}    

\begin{abstract} 
While the problem of grain alignment was posed more than 60 years ago the quantitative model of grain alignment that can account for the observed polarization arising from aligned grains has been formulated only recently. The quantitative predictions of the radiative torque mechanism, which is currently accepted as the dominant mechanism of grain alignment, open avenues to tracing magnetic fields in various astrophysical environments, including diffuse and dense interstellar gas, molecular clouds, circumstellar environments, accretion disks, comet tails, Zodiacal dust etc. At the same time, measurements of the absolute value of polarization and its variations can, in addition, provide unique information about the dust composition and dust environment. In the review I describe the analytical model describing well radiative torques acting on irregular grains and discuss how the alignment induced by radiative torques varies in the presence of superparamagnetic inclusions and pinwheel torques, e.g. arising from the H$_2$ formation over grain surface. I also describe observations that can establish whether  grains are superparamagnetic and whether recoils from H$_2$ formations are powerful enough to give rise to substantial uncompensated torques. Answering to these questions should allow for reliable modeling of astrophysical polarization with numerous important applications, from accounting for dust contribution in Cosmic Microwave Background polarization studies to obtaining magnetic field strength using Chandrasekhar-Fermi technique.
\end{abstract}


\section{Attempts to Explain Alignment}   

Polarization of starlight arising from aligned dust was discovered accidentally approximately 60 years ago (Hall 1949, Hiltner 1949). This gave rise to  attempts to explain the alignment. In the years that followed various interactions from paramagnetic relaxation (Davis \& Greenstein 1951) and streaming of grains (Gold 1951) to interaction with cosmic rays (Salpeter \& Wickramasinghe 1969) and photons (Harwit 1970) have been explored. Many key insights into the dynamics of grains are associated with Lyman Spitzer and Edward Purcell who addressed the problem of grain alignment on a number of occasions\footnote{My communications with Lyman Spitzer revealed to me his vision of grain alignment being one of the fundamental problems of interstellar medium, which solution is required for getting the quantitative insight into the role of magnetic fields.}. While we can refer the reader interested in the history of ideas on grain alignment to a review in Lazarian (2003), in this short publication we concentrate on the modern quantitative understanding of grain alignment.

Several mechanisms were proposed and elaborated to various degree (see Lazarian 2007 for a review), including the "textbook solution", namely, 
the paramagnetic Davis-Greenstein (1951) mechanism, which matured through intensive work since its introduction (e.g. Jones \& Spitzer 1967, Purcell 1979, Spitzer \& McGlynn 1979, Mathis 1986, Roberge et al. 1993, Lazarian 1997, Roberge \& Lazarian 1999). The mechanical stochastic alignment was pioneered by Gold (1951), who concluded that supersonic flows should align grains rotating thermally. Further advancement of the mechanical alignment mechanism (e.g. Lazarian 1994, 1995a) allowed one to extend the range of applicability of the mechanism, but left it as an auxiliary process, nevertheless. The major problem was that even the favorite alignment mechanism, the paramagnetic one, experienced severe problems explaining observational data. 

I feel that the attempts to solve the problem for spheres and spheroids, reminiscent of a theorist's favorite "spherical cow", were the major stumbling block for understanding of grain alignment. The first attempt to consider something which is not symmetric but has net helicity was a ground-breaking study by Dolginov \& Mytrophanov (1976). The authors considered there a grain that has different cross-sections for the extinction of the right- and left-polarized photons and predicted that such a grain was bound to spin up and get aligned when subjected to the anisotropic external radiative field. 

The study by Dolginov \& Mytrophanov (1976) had several deficiencies, however. First of all, it did not provide clear recipes about calculating the amplitude of radiative torques. In addition, the functional dependences of the torques calculated there were incorrect (Hoang \& Lazarian 2009). One way or another, the work  was mostly ignored for another  20 years until it attracted attention of Bruce Draine, who modified his publicly available DDSCAT code to calculate radiative torques acting on irregular grains.
This resulted in the explosion of interest to radiative torques. Empirical studies in Draine (1996), Draine \& Weingartner (1996, 1997), Weingartner \& Draine (2003) demonstrated that the magnitude of torques is substantial for irregular shapes studied. After that it became impossible to ignore the radiative torque alignment. Later, the spin-up of grains by radiative torques was demonstrated in laboratory conditions (Abbas et al. 2004).

The initial work on radiative torque alignment did not provide quantitative predictions for the grain alignment degree.  The multi-parameter space presented by grain alignment induced by radiative torques (henceforth RATs) posed an insurmountable  problem for the "brute force" numerical approach. At the same time, both the interpretation and modeling of polarization call for simple recipes to parameterize effects of grain alignment. This is not feasible with numerical calculations which suggest that RATs depend on grain shape, grain size, radiation spectrum, grain composition, and the angle between the radiation direction and the magnetic field. Consequently, the important empirical studies above had limited predictive powers and were used to demonstrated the radiative torque effects sometimes using one grain shape, one grain size, one wavelength of light, and one direction of the light beam with respect to the magnetic field. 

The quantitative stage of radiative torque studies required theoretical models describing radiative torques. In Lazarian \& Hoang (2007a) we proposed a simple model of RATs which allowed a good analytical description of the alignment. This model was elaborated and extended in Lazarian \& Hoang (2008) and Hoang \& Lazarian (2008, 2009ab).   

Recent reviews on grain alignment include an extended one by Lazarian (2007). However, the subject of grain alignment has been developing so rapidly, that it does not
reflect all the key present-day ideas. A short review by Lazarian \& Hoang (2009) concentrates on the recent developments in the field. However, both reviews are focused on how understanding of grain alignment improves magnetic field tracing. In the paper below we consider, in addition, how quantitative understanding of grain alignment can shed light onto grain composition and grain environment. I would like to stress that grain alignment is a genetic property of astrophysical grains which applies not only to dust in interstellar gas and molecular clouds, but also to dust in accretion disks, AGN environments, circumstellar regions, solar system etc. 
 
\section{Analytical Model for Radiative Torques} 

\begin{figure}
\includegraphics[width=0.7\textwidth,angle=270]{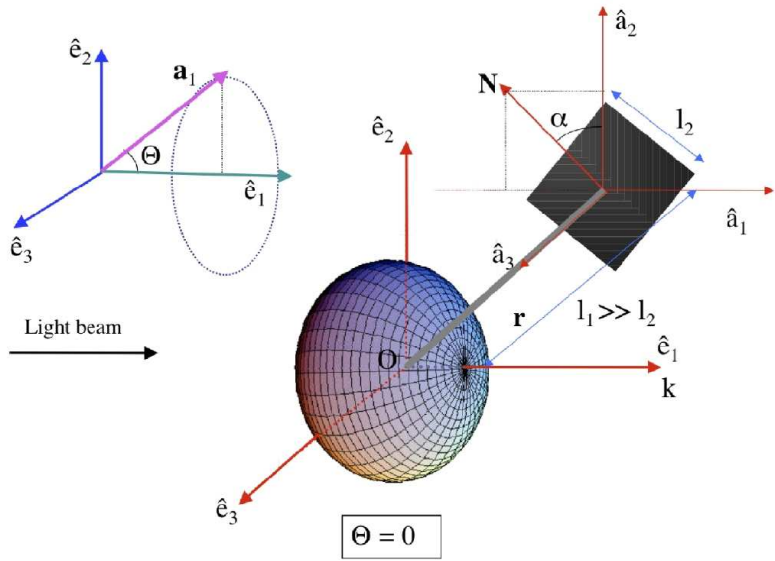}
\caption{
\small
A model of a ``helical'' grain,
that consists of a spheroidal body with a mirror at an angle $\alpha$ attached to it ($\alpha$ is chosen to be $\pi/4$ in the standard LH07 model).
 The ``scattering coordinate system'' which
illustrates the definition of torque components: ${\bf a}_1$ is directed
along the maximal inertia axis of the grain; ${\bf k}$ is the direction of radiation.
The projections of normalized radiative torques $Q_{e1}$,
$Q_{e2}$ and $Q_{e3}$ are calculated in this reference frame. From Lazarian \& Hoang 2007a.}
\label{AMO1}
\end{figure}
Dolginov \& Mytrophanov (1976) were the first to propose an analytical model to describe RATs. However, their analytical predictions for the model of a twisted grain obtained using Raylegh-Hans approximation are in error, as they are inconsistent with the DDSCAT numerical simulations (see Hoang \& Lazarian 2009b). 

Lazarian \& Hoang (2007a, henceforth LH07) assumed that the most important property is grain helicity and this property should carry over to a geometric optics limit. 
Therefore they suggested a simple model of a helical grain that was shown by DDSCAT calculations to reproduce well the essential properties of RATs acting on
irregular grains. The model is shown in Fig.~\ref{AMO1} and it consists of an ellipsoidal body with a mirror attached to
its side. LH07 termed the model AMO, which is an abbreviation from Analytical MOdel. Note,  that grains can be both  
``left-handed'' and  ``right-handed''. For our grain model to become ``right handed'' the mirror should be turned  by 90 degrees. 
Our studies with DDSCAT confirmed that actual irregular grains also vary in handedness. A substantial difference in the RATs acting on
right and left handed irregular grains was a source of earlier confusion.

 To describe torques acting of AMO and irregular grains LH07 chose a system of reference with the direction of light along a vector ${\bf e_1}$, and the grain axis of maximal inertia moment ${\bf a}_1$ being in the ${\bf e_1}$, ${\bf e_2}$ plane (see Figure\ref{AMO1}). The latter condition ensures that the torques calculated in the $e_1-e_3$ reference frame do not change as the grain precesses around ${\bf k}$. If we recall that the equation for the changes of the angular momentum for a top is
$d{\bf J}/dt={\bf Q}$, it is easy to see that angle $\Theta$ and angular velocity of AMO depends only on the torque components $Q_{e1}$ and $Q_{e2}$. The third component $Q_{e3}$ induces grain precession only. In the absence of magnetic field, this would induce the direction of the beam to serve as the alignment direction, but in most cases the precession induced by $Q_{e3}$ is subdominant to the Larmor precession induced by the ambient, e.g. interstellar, magnetic field (see more discussion in LH07). Interestingly enough, the conclusion of $Q_{e3}$ is not important in terms of the RAT alignment is also true for the presence of thermal fluctuation (see Hoang \& Lazarian 2008) and inefficient internal relaxation (see Hoang \& Lazarian 2009b) when the alignment of angular momentum and axis ${\bf a}_1$ is not enforced\footnote{It was shown in LH07 that the only component of RATs present for an ellipsoidal grain is $Q_{e3}$. Naturally, this component cannot produce the RAT alignment, as the helicity of an ellipsoidal grain is zero.}. This observation allowed LH07 to simplify the problem and consider only two torque components, namely, $Q_{e1}$ and $Q_{e2}$ instead of three RAT components.

\begin{figure}
\includegraphics[width=0.6\textwidth,angle=270]{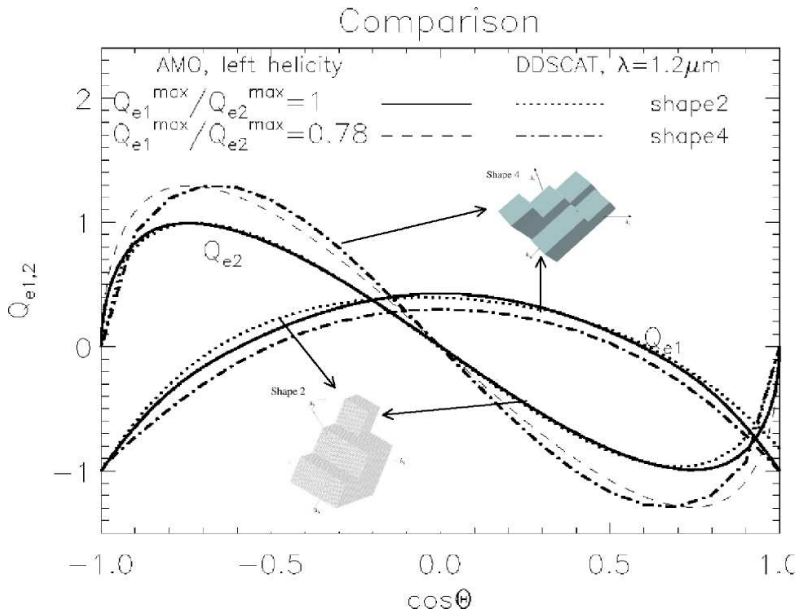}
\caption{
\small The variations of the torques with the angle between the line of sight and the axis of maximal moment of inertia $\Theta$ for AMO and irregular grains. For the latter RATs were obtained via DDSCAT calculations. Images of the irregular grains, namely, shapes 2 and 4 are also shown in the figure. Both irregular grains and AMO have left helicity. From Lazarian \& Hoang 2007a.}
\label{AMO2}
\end{figure}

\begin{figure}
\begin{center}
\includegraphics[width=0.8\textwidth]{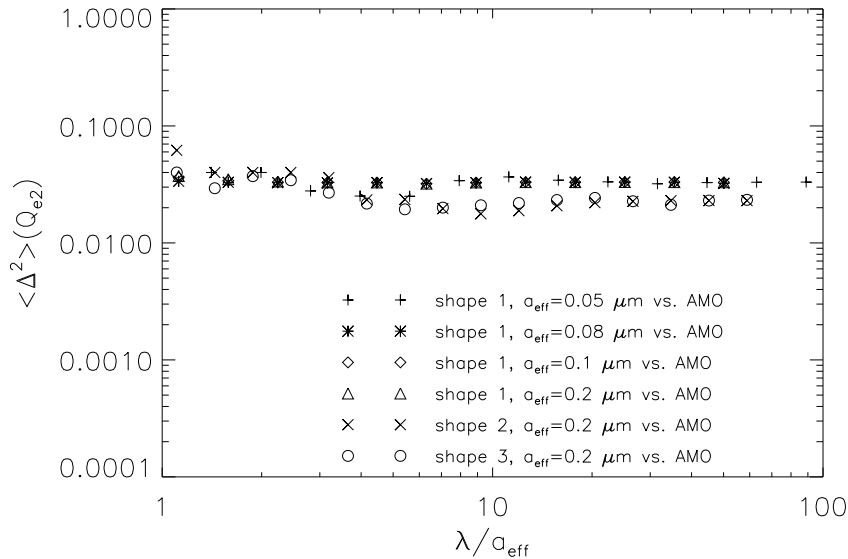}
\end{center}
\caption{\small Numerical comparison of the torques calculated with DDSCAT for irregular grains for different wavelength and the analytical model (AMO) of a helical grain. The quantity
$\langle \Delta^2 \rangle$ is defined by Eq.~(\ref{chi_eq}).  From Lazarian \& Hoang 2007a. }
\label{chi}
\end{figure}

The functional dependences of torques $Q_{e1}(\Theta)$ and $Q_{e2}(\Theta)$, where $\Theta$ is an angle between the axis ${\bf a}_{1}$ and the radiation direction, were shown to be very similar for the analytical model in Fig.~\ref{AMO1} and irregular grains subject to radiation of different wavelengths. 
In Figure~\ref{AMO2} this correspondence is shown for two irregular grains (Shape 2 and Shape 4 in LH07) and AMO. This remarkable correspondence is further quantitatively illustrated in Fig.~\ref{chi} using a function:

\begin{equation}
\langle \Delta^{2}\rangle(Q_{e2})=\frac{1}{\pi (Q_{e2}^{max})^{2}} \int^{\pi}_{0} [Q_{e2}^{irregular}(\Theta) -Q_{e2}^{model} (\Theta)]^{2} d\Theta,
\label{chi_eq}
\end{equation}
which characterizes the deviation of the torques $Q_{e2}$ calculated numerically for irregular grains from the analytical prediction in the LH07 model. 

While the functional dependence of torque components $Q_{e1}(\Theta)$ and $Q_{e2}(\Theta)$ coincides for grains of various shapes, their amplitudes vary for different  grains and different radiation wavelengths. In fact, LH07 showed that the radiative torque alignment can be fully determined if  
the ratio $q^{max}=Q_{e1}^{max}/Q_{e2}^{max}$ is known. In terms of practical calculations, this {\it enormously} simplifies the calculations
of radiative torques: instead of calculating two {\it functions} $Q_{e1}(\Theta)$ and $Q_{e2}(\Theta)$ it is enough to calculate just two {\it values} $Q_{e1}^{max}$ and $Q_{e2}^{max}$. According to LH07 the maximal value of the function $Q_{e1}(\Theta)$ is achieved for $\Theta=0$ of the function $Q_{e2}(\Theta)$ is achieved at $\Theta=\pi/4$. In other words, one can use a {\it single number} $q^{max}=Q_{e1}^{max}/Q_{e2}^{max}=Q_{e1}(0)/Q_{e2}(\pi/4)$ instead of
{\it two functions} to characterize grain alignment. Thus, it is possible to claim that the $q^{max}$-ratio is as important for the alignment as the grain axis
ratio for producing polarized radiation by aligned grains. 

\begin{figure}
\includegraphics[width=0.8\textwidth]{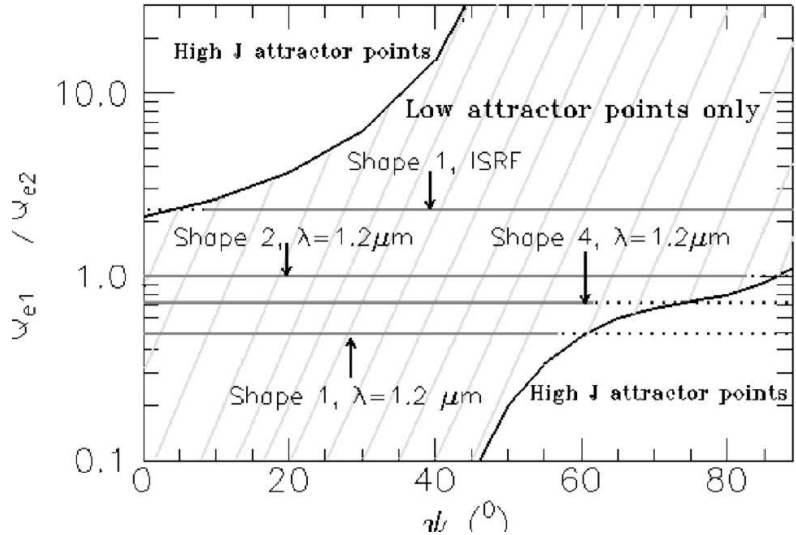}
\caption{\small  Parameter space for which grains have only low-$J$ attractor point and both low-$J$ and high attractor point. $\psi$ is the angle between the direction towards a point radiation source and magnetic field. In the situation when the high-$J$ attractor point
is present grains eventually get there and demonstrate perfect alignment. In the situation when only low-$J$ attractor point is present, the alignment is partial. From Lazarian \& Hoang 2007a. }
\label{space}
\end{figure}

Studying the RAT alignment LH07 corrected the treatment of grain dynamics in Draine \& Weingartner (1997). As a result, instead of cyclic trajectories in the latter paper, LH07 reported the situation when, instead of accelerating grains, RATs were slowing grains down. This slowing down was previously reported in Weingartner \& Draine(2003), but erroneously attributed to the effect of thermal fluctuations. On the contrary, LH07 had the same set up as in Draine \& Weingartner (1997), i.e. without any thermal fluctuations. Therefore, the effect of braking grain rotation by RATs, irrespectively of any other factors, was established. This effect happened to be very important for the alignment\footnote{This slowing down of grains in
in the presence of thermal fluctuations (see Hoang \& Lazarian 2008) does not bring grains to a complete stop, but results in creating of low-$J$ attractor point, in agreement with an earlier empirical study in Weingartner \& Draine (2003).}. Thus, apart from high attractor points, LH07 found that for a range of
$q^{max}$ and the angle $\psi$ between the radiation beam and magnetic field direction only low-$J$ attractor points exist (see Figure~\ref{space}). A later study by Hoang \& Lazarian (2008) established that when low-$J$ and high-$J$ attractor points coexist, the high-$J$ points are more stable and therefore an external stochastic driving, e.g.
arising from gaseous bombardment, brings grains to high-$J$ attractor points. This transfer can take several damping times which may have observational consequences for the alignment in the presence of varying sources of radiation, e.g. supernovae. However, for the steady-state interstellar alignment for the parameter space for which a high-$J$ attractor point exists (see Figure~\ref{space}), one can safely assume that grains are aligned with high-$J$. As the degree alignment at high-$J$ attractor point is higher than at a low-$J$ attractor point, this is a very peculiar effect of improving alignment through random gaseous bombardment! 

When does the alignment happens at low-$J$ attractor points? Figure~\ref{space} shows predictions for the existence of low-$J$ and high-$J$ attractor points for the analytical model (AMO) for the parameter space given by $q^{max}$ and the angle $\psi$. Individual horizontal lines correspond to particular grain shapes with a given $q^{max}$. For the interstellar radiation field (ISRF) the calculations of $q^{max}$ are performed using torques averaged over the interstellar spectrum of wavelengths (see LH07 for more details). We see that the correspondence in terms of predicting the distribution of high-$J$ and low-$J$ attractor points is
also good, which, however, is not surprising due to the good correspondence between the functional dependences obtained for the AMO and irregular grains depicted in Figure~\ref{AMO1}.
Further research in Hoang \& Lazarian (2009b) revealed that when the radiation arises not from a point source but the radiation field has a complex structure defining parameter spaces for low-$J$ and high-$J$ attractor points is possible if the radiation field is decomposed into multipoles, e.g. dipole and quadrupole.  
\begin{figure}
\includegraphics[width=0.8\textwidth]{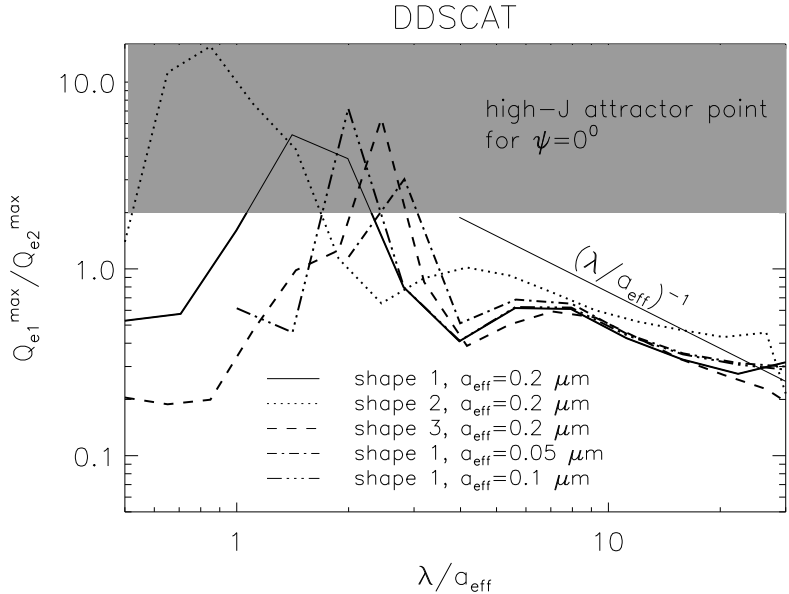}
\caption{\small  The magnitude of the ratio $q=Q_{e1}^{max}/Q_{e2}^{max}$ that characterize the radiative torque alignment of grains depends on both grain shape and the wavelength of radiation. From Lazarian \& Hoang 2007a. }
\label{wavelength1}
\end{figure} 

The analytical description of RATs in AMO does substantially simplify modeling of the RAT alignment. However, modeling of the polarization arising from grains with the infinite variety of shapes and for various spectra may still present a problem. The approaches to handling this problem are summarized below.

As the torques $Q_{e1}$ and $Q_{e2}$ vary with the wavelength this results in the change of $q^{max}$ (see Figure~\ref{wavelength1}). The figure reveals systematic changes of $q^{max}$
that can be used in future for developing models of polarization for interstellar grains. 
 
The magnitude of the radiative torques also changes in a systematic way as shown in 
Figure~\ref{wavelength2}. This fact was already used to simplify calculations of grain alignment 
in T-Tauri disks in Cho \& Lazarian (2007). Note that the calculations for low $\lambda/a_{eff}$ get not reliable with DDSCAT code that was employed. Ray tracing techniques would be much more appropriate for such calculations. Large grains are known to  exist in circumstellar accretion disks and their alignment is important. 

Our studies of $q^{max}$ are limited to a handful grain shapes. To make modeling more reliable it is important to perform more DDSCAT studies of more grain shapes in order to reveal the statistics of $q^{max}$. Naturally, obtaining the statistics of $q^{max}$ is a much more simple task compared to dealing with the multitude of functions that can be obtained via numerical calculations of radiative torques.

\begin{figure}
\includegraphics[width=0.8\textwidth]{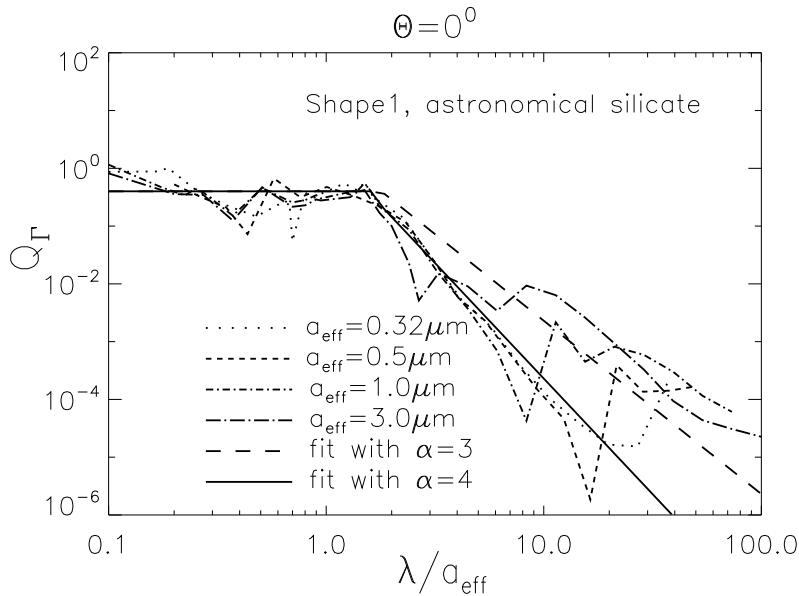}
\caption{\small Normalized torques for grains of different sizes and wavelengths. The torque 
amplitude is proportional to radiation intensity. The most efficinet alignment is for grains larger than $\lambda/2$. However, the alignment of grains substantially smaller than  the radiation wavelength can also be present provided that the radiation is strong enough. From Lazarian \& Hoang 2007a. }
\label{wavelength}
\end{figure} 

\section{Grains get aligned with long axes perpendicular to ${\bf B}$ without paramagnetic relaxation}
\label{s:why}

Observations testify that interstellar grains tend to get aligned with long axes perpendicular to the ambient magnetic field, the fact that was frequently used to argue that the Davis-Greenstein (1951) mechanism is responsible for the alignment. Can RATs explain this observational fact? We note, that other mechanisms, e.g. mechanical alignment of oblate and prolate grains, i.e. Gold (1951) alignment, align grains both perpendicular and parallel to magnetic field, depending on the direction of the gaseous flow in respect to magnetic field. 

The direction of alignment was not an issue in the original Draine \& Weingartner (1996) study, as there the alignment was assumed to be paramagnetic, while the radiative torques were associated only with the grain spin-up. As we discussed above, this is not what is going on and RATs do produce the alignment of their own. Using the AMO, one can show that the RATs acting on their own, without any effect of paramagnetic relaxation, tend to align grains the "right way", i.e. in agreement with observations. The exception is a narrow range of angles when the light beam direction is nearly perpendicular to the direction of magnetic field. In the latter situation the alignment could be "wrong", i.e. with long grain axes parallel to magnetic field. As the analysis testifies that the "wrong" alignment happens with low angular momentum, the thermal fluctuations cause wobbling of grain axes about grain angular momentum (Lazarian 1994, Lazarian \& Roberge 1997). The angular amplitude of this wobbling for any reasonable grain temperature, e.g. $T_{grain}=10K$ , exceeds the range of the beam angles for which the alignment is "wrong" (LH07, Hoang \& Lazarian 2008).  Thus a remarkable fact emerges: grains get {\it always} aligned with long axes perpendicular to ${\bf B}$!

Is it possible to understand this on a more intuitive level? Below we try to provide an explanation why the situation when ${\bf J}$ is aligned with magnetic field ${\bf B}$ is
special. For the sake of simplicity, we disregard grain wobbling and assume that due to internal relaxation of energy ${\bf J}$ is perfectly aligned with the axis of maximal inertia.
 Therefore, it is sufficient to follow the dynamics of angular momentum to determine grain axes alignment. 
\begin{figure}[h]
\includegraphics[width=0.5\textwidth, angle=270]{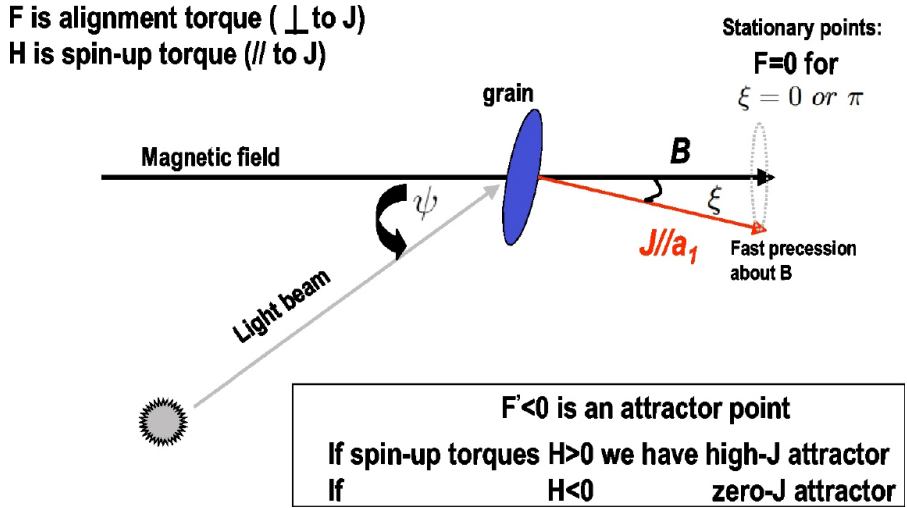}
\caption{\small A simplified explanation of the grain alignment by radiative torques. The grain, which is depicted as a spheroid in the figure, in fact, should be irregular to experience non-zero radiative torque. 
The positions ${\bf J}$ parallel (or anti-parallel) to ${\bf B}$ correspond to the stationary points as at these positions the component of torques that changes the alignment angle vanishes. As 
internal relaxation makes ${\bf J}$ aligned with the axis $a_1$ of the maximal moment of grain inertia, the grain gets aligned with long axes perpendicular to ${\bf B}$.}
\label{alignment}
\end{figure}

As we discussed earlier, only two components of the torque, namely $Q_{e1}$ and $Q_{e2}$ are important for the processes of alignment. The choice of scattering frame axes
and the perfect alignment of ${\bf J}$ with the axes of maximal moment of inertia ensures that ${\bf J}$ is in the $e_1 e_2$ plane. Torques $Q_{e1}$ and $Q_{e2}$ change both
the direction and the amplitude of ${\bf J}$. In this situation it is convenient to present the action of these torques in the reference system with one of the axes directed along ${\bf J}$ and the other axis being perpendicular to ${\bf J}$ (still in the $e_1 e_2$ plane).  
Let us denote the component of torque parallel to ${\bf J}$ by ${\bf H}$ and perpendicular to ${\bf J}$ by ${\bf F}$. It is evident that ${\bf H}$ torque spins the grain up or down and
the torque ${\bf F}$ aligns the grain.

Figure~\ref{alignment} illustrates the process of alignment. The angular momentum ${\bf J}$ is precessing about magnetic field ${\bf B}$ due to the magnetic moment\footnote{Magnetic moment arises from the Barnett effect as the grain rotates (Dolginov \& Mytrophanov 1976).} of a grain. The alignment torques ${\bf F}$ are perpendicular to ${\bf J}$ and therefore as  ${\bf J}$ gets parallel to ${\bf B}$ the fast precession of the grain makes the torques averaged over ${\bf J}$ precession vanish as $\xi\rightarrow 0$. Thus the positions corresponding to ${\bf J}$ aligned with ${\bf B}$ are stationary points, irrespectively of the functional forms of radiative torques, i.e. of components $Q_{e1}(\Theta)$ and $Q_{e2}(\Theta)$. In other words, grain do not experience aligning torques when $\xi=0$ or $\pi$. Whether these stationary points are stable, i.e. attractor points, or unstable, i.e. repellor points, it is impossible to say unless the functional form of torques is given. Thus, on the intuitive level, one can understand why grains may get {\it perfectly} aligned with ${\bf J}$ parallel to
${\bf B}$, but quantitative predictions of the alignment degree can be only obtained with the help of AMO.

\section{Important Special Cases of Alignment}

Our earlier discussion was focused on ordinary paramagnetic grains, for which the effect of paramagnetic relaxation is absolutely negligible compared to the effect of RATs. We also neglected the possible action of pinwheel torques, which may arise, for instance, from H$_2$ formation over grain catalytic sites, as it is discussed by Purcell (1979). In addition, we assumed that internal relaxation of energy is sufficiently strong to induce the alignment of the grain axis of maximal moment of inertia with ${\bf J}$. These were the simplifying assumptions of LH07 study. Below we consider special situations when one of these conditions is not true. Future observations should determine how special or generic the superparamagnetic response of the grains and the existence of strong pinwheel torques.

\subsection{Radiative torque alignment of superparamagnetic grains}
\label{magnetic}

Superparamagnetic grains, i.e. grains with enhanced paramagnetic relaxation, 
 were invoked by Jones \& Spitzer (1967) within the model of paramagnetic alignment (see also Mathis 1986, Martin 1995, Goodman \& Whittet 1994, Roberge \& Lazarian 1999). 
What does happen when the dynamics of grains is determined by RATs? We see from Fig.~\ref{space} that 
for a substantial part of the parameter space grains are driven to the low-$J$ states, i.e. {\it subthermally}, which is in contrast to a widely spread belief 
 that in the presence of RATs most of the interstellar grains must rotate at  $T_{rot}\gg T_{gas}$.
\begin{figure}
\includegraphics[width=0.7\textwidth]{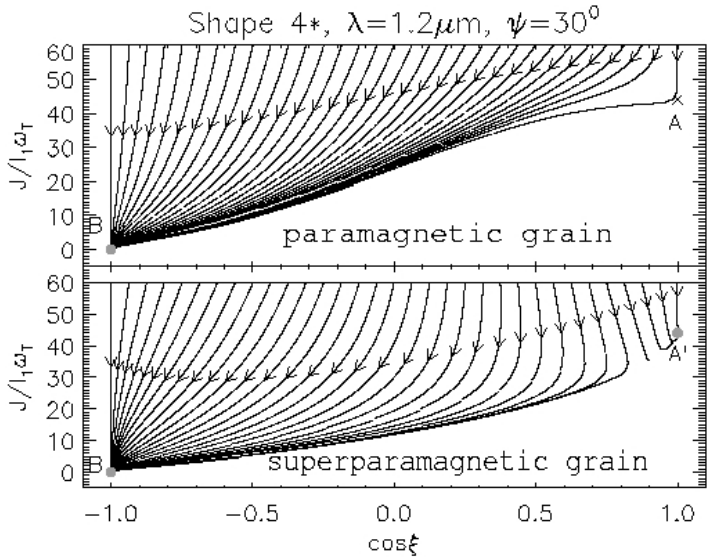}
\caption{\small RAT alignment of superparamagnetic grains. A paramagnetic grain gets only a low-$J$ attractor point. For the same set of parameters a superparamagnetic grain gets also a high-$J$ attractor point. High-$J$ attractor points are more stable than the low-$J$ attractor points. As a result, all grains eventually end up
at the high-$J$ attractor point. From Lazarian \& Hoang (2008).}
\label{superparamagnetic}
\end{figure} 

The picture above, however, is different if superparamagnetic grains dissipate (via Davis-Greenstein process) rotational energy on the time scales shorter
than the gaseous damping time.  Lazarian \& Hoang (2008)  found that  trajectory maps of grains with superparamagnetic inclusions {\it always} exhibit 
high-$J$ attractor points. 

Fig.~\ref{superparamagnetic} shows that for superparamagnetic grains subject to a diffuse interstellar radiation field most grains still get
to the low attractor point. As the high-$J$ attractor point is more stable compared to the low-$J$ attractor point
superparamagnetic grains get transfered by gaseous collisions from the low-$J$ to high-$J$ attractor points. Thus, superparamagnetic grains 
always {\it rotate at high rate} in the presence of RATs and their alignment is {\it perfect}. 

\subsection{Radiative torque alignment in the presence of pinwheel torques}
\label{pinwheel}

Pinwheel torques were considered by Purcell (1979) in the context of paramagnetic alignment. For instance, the action of
active sites forming  $H_2$ molecules can be similar to the action of tiny rocket engines spinning up the grain.

\begin{figure}
\includegraphics[width=0.7\textwidth]{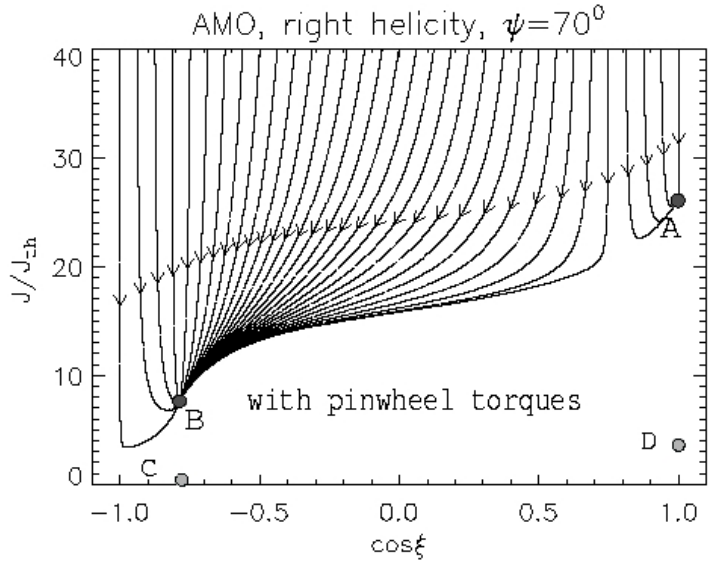}
\caption{\small  Grain alignment by radiative torques in the presence
of  pinwheel torques. The shown case corresponds to the presence of both the low-$J$ and high-$J$ attractor points in the absence of pinwheel torques. In the case when only a low-$J$ attractor
point exists the strong pinwheel torques lift the low-$J$ attractor point enhancing the alignment. From Hoang \& Lazarian 2009a.}
\label{suprathermal}
\end{figure}

How do these torques also affect the RAT alignment? Hoang \& Lazarian (2009a) showed
that the sufficiently strong pinwheel torques can create new high-$J$ attractor points (see Figure~\ref{suprathermal}). Therefore for
 strong pinwheel torques  arising from H$_2$ formation, one may observe the correlation
of higher degree of polarization with the atomic hydrogen content in the media, provided that H$_2$ torques as strong
as in Purcell (1979) and the subsequent papers (see Spitzer \& McGlynn 1979, Lazarian 1995, Lazarian \& Draine 1997). 
The implicit assumption for observing this correlation is, however, that the grains are not superparamagnetic. For superparamagnetic grains
the alignment, as we discussed above, is perfect anyhow. 

\subsection{Alignment with negligible internal relaxation}

It may be shown (see Lazarian \& Hoang 2009) that for sufficiently large grains the internal relaxation over the time-scale of RAT alignment is negligible. Large grains, however, are present e.g. in accretion disks and comets. The polarization from accretion disks can provide one with an important insight into magnetic of the objects (see Cho \& Lazarian 2007), while understanding of the alignment for comet dust is important for explaining circular polarization observed (see below). Thus,
a proper description of RAT alignment in the absence of internal relaxation is important. 

In the absence of internal relaxation grains can get aligned not only with long axes perpendicular to the angular momentum ${\bf J}$, but 
also with the longest axis parallel to ${\bf J}$, i.e. the axis of minimal moment of grain inertia parallel to ${\bf J}$. This complicates the analysis compared to the case of interstellar grains, for which the internal relaxation is very fast. The corresponding problem was addressed in Hoang \& Lazarian (2009b, henceforth HL09). The results of the latter study are summarized in Table~1.
{\small
\begin{table}
\caption{Attractor points with and without internal relaxation for AMO. "L" denotes "long axes of the grain".}
\begin{tabular}{llcll} \hline\hline\\
\multicolumn{1}{c}{\it Without relaxation (HL09)} & & \multicolumn{1}{c}{\it With relaxation (LH07)}\\[1mm]
\hline\\
{{\bf High-J }}& {{\bf Low-J }} &{{\bf High-J}}&{{\bf Low-J}}\\[1mm]
&&&&\\[1mm]
{$\bJ\|\bB$}& {$\bJ$ $\|$ or at} &{$\bJ\| \bB$}&{$\bJ$  $\|$ or at}\\[1mm]
&{angle with $\bB$}&&{angle with $\bB$}\\[1mm]
{L $\perp \bB$}& {L $\perp$ or $\|$ $\bJ$} &{L $\perp \bB$}&{L $\perp \bB$}\\[1mm]\\[1mm]
\hline\hline\\
\end{tabular}
\end{table}
}

The study of HL09 is suggestive that grains do preferentially get aligned with long axes perpendicular to magnetic field even without internal relaxation. Indeed,
this is consistent with the finding there that such alignment happens for the high-$J$ attractor points of AMO. If this is the case, when the low-$J$ and high-$J$ attractor points coexist the steady-state alignment will happen only with high-$J$ attractor point (see discussion in \S\ref{magnetic}).  The "wrong" alignment happens only with low-$J$ attractor points and it is suggestive that in the presence of gaseous bombardment the grains may still spend more time in the vicinity of the high-$J$ repellor point, as was shown in numerical simulations by Hoang \& Lazarian (2008). The conclusions obtained with AMO are consistent with a limited parameter study obtained in HL09 for an irregular grain. However, it is clear that more extensive studies of the RAT alignment in the absence of internal relaxation are required. These studies are of practical interest for modeling alignment in circumstellar accretion disks, where large grains with slow rates of internal are known to be present.

\section{Getting insight into grain composition and grain environment}  

The predictions that the alignment is always perfect for superparamagnetic grains opens ways to testing this hypothesis using polarimetry. For instance, the variations of the 
polarization with the angle between the direction of light beam and magnetic field in the situations that RATs are strong would reveal that grains are not superparamagnetic. Similarly, variations of the degree of grain alignment with fraction of atomic hydrogen in their environments would testify that the Purcell's pinwheel torques arising from
H$_2$ formation strong and, simultaneously, that grains are not superparamagnetic.

On the contrary, statistical studies indicating that the variations of the polarization arise only from the structure and the orientation of magnetic field in respect to the line of sight, while the grains are always perfectly aligned, would testify that grains are superparamagnetic. The structure and orientation of magnetic field can be tested by the technique proposed in Falceta-Gonzalvez et al. (2008). Additional information on the orientation of 3D vector of magnetic field can be obtained using a new technique of atomic alignment polarimetry (Yan \& Lazarian 2006, 2007, 2008). Note, that by increasing the sample of observations or using the observations when the relative orientation of magnetic field to the sources of orientation is known, e.g. for surcumstellar regions, it is possible to place stringent constraints on the degree of grain alignment. 

While the confirmation of the evidence of the strength of H$_2$ formation torques may give insight into the intimate details of the formation of molecular hydrogen, the consequences of 
testing of supermagnetic nature of interstellar grains are much broader. First of all, if grains are shown to be superparamagnetic, the modeling of polarization from them can be much simplified as assuming perfect alignment whenever the RATs are strong enough will be possible (see Cho \& Lazarian 2005, 2007, Pelkoen et al. 2007, Bethell et al. 2007). Moreover, superparamagnetic grains emit polarized magneto-dipole emission in the microwave range (Draine \& Lazarian 1999), which can interfere with the attempts to measure Cosmic Microwave Background polarization. 

One should keep in mind that if proven that interstellar grains are not superparamagnetic this does  not preclude grains in other environments to be superparamagnetic. Future research should make use  of the predictions of the quantitative theory of grain alignment to study not only magnetic fields, but also the environment and the composition of grains.


\acknowledgements 
Our research was supported by the NSF Center for Magnetic Self-Organization in Astrophysical and Laboratory Plasmas and the NSF grant 0507164.


\end{document}